\documentstyle[sprocl,psfig]{article}

\def\beq{\begin{equation}}
\def\eeq{\end{equation}}
\def\bea{\begin{eqnarray}}
\def\eea{\end{eqnarray}}

\begin{document}
\renewcommand{\thefootnote}{\fnsymbol{footnote}}

\title{\hfill{\small {\bf MKPH-T-96-11}}\\
PROBING HADRON STRUCTURE BY REAL AND VIRTUAL PHOTONS
\footnote{Dedicated to Walter Greiner on the occasion of his 60th birthday}}

\author{ HARTMUTH ARENH\"OVEL }

\address{Johannes Gutenberg-Universit\"at, D-55099 Mainz, Germany}

\maketitle
\abstracts{
The importance of subnuclear degrees of freedom in photon absorption and 
electron scattering by nuclei is discussed. 
After a short introduction into the basic
properties of one-photon processes and a very brief survey on the nuclear 
response in the various regions of energy and momentum transfers, the 
particular role of subnuclear degrees of freedom 
in terms of meson and isobar degrees of freedom 
is considered. Their importance 
is illustrated by several reactions on the deuteron which are currently 
under study at c.w.\ electron machines like MAMI in Mainz. 
}

\section{Introduction}
\label{intro}

Since the beginning of nuclear physics, when the existence of the atomic 
nucleus was deduced by Rutherford from the famous experiments of 
$\alpha$-particle scattering on a gold foil  
by Geiger and Marsden
up to the deep inelastic scattering 
experiments at Stanford by Friedman, Kendall and Taylor
revealing the parton substructure of nucleons, 
the electromagnetic probe has always played an outstanding role in the 
study of nucleon and nuclear structure~\cite{Are94}. 

One of the major goals of present day research in this field is to clarify 
the role of subnuclear degrees of freedom (d.o.f.) in the structure of 
nuclei as well as their relation to the underlying quark-gluon dynamics of 
QCD. 
In this talk I will concentrate on the manifestation of 
of meson and isobar degrees of freedom 
in electromagnetic processes where they contribute as two-body meson 
exchange currents (MEC) and via nuclear isobar configurations (IC). 
The latter are nuclear wave function components containing internally excited 
nucleons (isobars). 

\section{Properties of the Electromagnetic Probe}
\label{propem}
The special role of the electromagnetic inter\-action in unravelling the
micro\-struc\-ture of the world is due to the fact that 
(i) its properties as a classical field are well known, 
(ii) the electro\-magnetic inter\-action ful\-fills the basic require\-ments 
of a fundamental inter\-action incorpo\-rating rela\-tivity and represen\-ting 
the simplest case of a gauge theory, and 
(iii) the electro\-magnetic inter\-action is weak enough to allow
lowest order pertur\-bation treat\-ment resul\-ting in simple and unique
inter\-pretations of ex\-perimental results.
However, this weakness constitutes also a disadvantage since 
the cross sections for photoreactions and electron scattering are
considerably smaller than for pure hadronic reactions.

The lowest order processes are the one-photon processes like photon
absorption (or emission) and electron scattering which is governed by the
exchange of one virtual photon. 
The main differences between photoabsorption and electron scattering
or real and virtual photon processes, respectively, are:

(i) For real photons one has a fixed relation between energy and momentum
transfer ($\vec q^{\,2}=\omega^2$) while for the exchange of a virtual 
photon in electron scattering the four momentum is space-like
($\vec q^{\,2}\ge \omega^2$) allowing an independent variation of energy
and momentum transfer within the spacelike region.

(ii) Real photons have only transverse polarization so that only the transverse
current density contributes while virtual photons have both, transverse and
longitudinal polarization allowing also the charge density to contribute.

Qualitatively one may distinguish different regimes of the nuclear response
to photo absorption. At low energies below particle emission threshold, one
finds sharp resonances corresponding to the excitation of bound excited
nuclear states. Above particle emission threshold for photon energies of
about 10 to 30 MeV the dominant feature is
the giant dipole resonance. It is a collective nuclear mode which exhausts
almost one classical TRK-sum rule. It can be
described in a phenomenological two-fluid model as an oscillation of the
proton against the neutron fluid or in a microscopic description as a
coherent superposition of one-particle one-hole excitations. Increasing the 
energy, one enters the domain of
short-range correlations or the so-called quasi-deuteron region, where the
leading process is the absorption by a correlated
two-nucleon pair emitted mainly back to back. Here, the major 
contributions come from MEC. 
Above pion production threshold, the dominant mode of absorption is the isobar
excitation of a nucleon to a $\Delta$ or to higher nucleon 
resonances.

In electron scattering one can explore the same dynamical regions but with 
the additional possibiliy of independent variation of the momentum transfer.
It allows, for example, in the elastic process ``cum grano salis''
the determination of ground state charge and magnetization densities.

\section{Subnuclear Degrees of Freedom in Deuteron Break-up}
\label{subdof}
Now I will turn to the discussion of subnuclear d.o.f.\ in 
deuteron disintegration by photons and electrons. 
These subnuclear d.o.f.\ are described in terms of meson exchange currents 
(MEC) and isobar configurations or currents (IC). 
The special role of the electromagnetic deuteron break-up is
a consequence of 
(a) the simple structure of the two-body system allowing exact
solutions at least in the nonrelativistic domain, and
(b) the specific features of the electromagnetic
probe as discussed in the preceding section. The continued interest in
this process has persisted over more than sixty years because
the two-body system is in fact a unique laboratory, in particular for the
study of subnuclear d.o.f.~\cite{ArS 91}. 
One may summarize the evidence for MEC and IC 
in $\gamma^{(\ast)}+d\rightarrow p+n$ as follows: \\
(i) The strongest MEC contributions appear in $E1$ in $d(\gamma,N)N$, but 
they are mostly covered by the Siegert operator. \\
(ii) A clear signature of MEC is furthermore observed in $M1$ in $d(e,e')np$ 
near break-up threshold at higher momentum transfers. \\
(iii) In the $\Delta$-region one finds a strong manifestation of IC. 


\subsection{MEC and relativistic effects in electrodisintegration}\label{eldeut}

\noindent
\begin{minipage}[t]{5.6cm}
\centerline{\psfig{figure=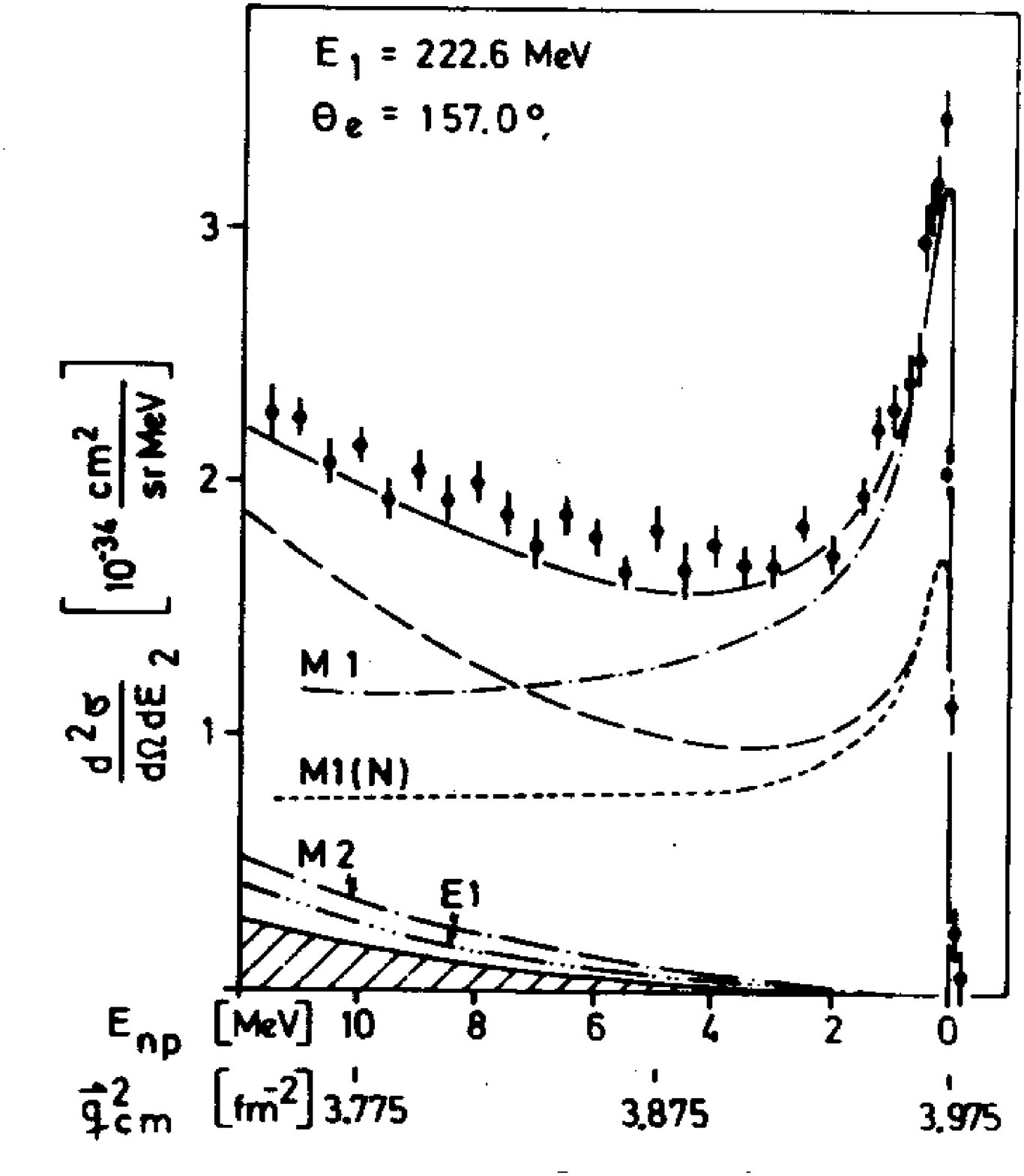,width=5.6cm}}
\noindent
{\small Fig.\ 1: The transverse deuteron form factor compared to calculations 
with the Hamada-Johnston potential for normal (N) theory
and additional MEC and IC (T) (from~\cite{SiB79}).}
\label{fig:th1}
\end{minipage}\hfill
\begin{minipage}[t]{5.5cm}
\centerline{\psfig{figure=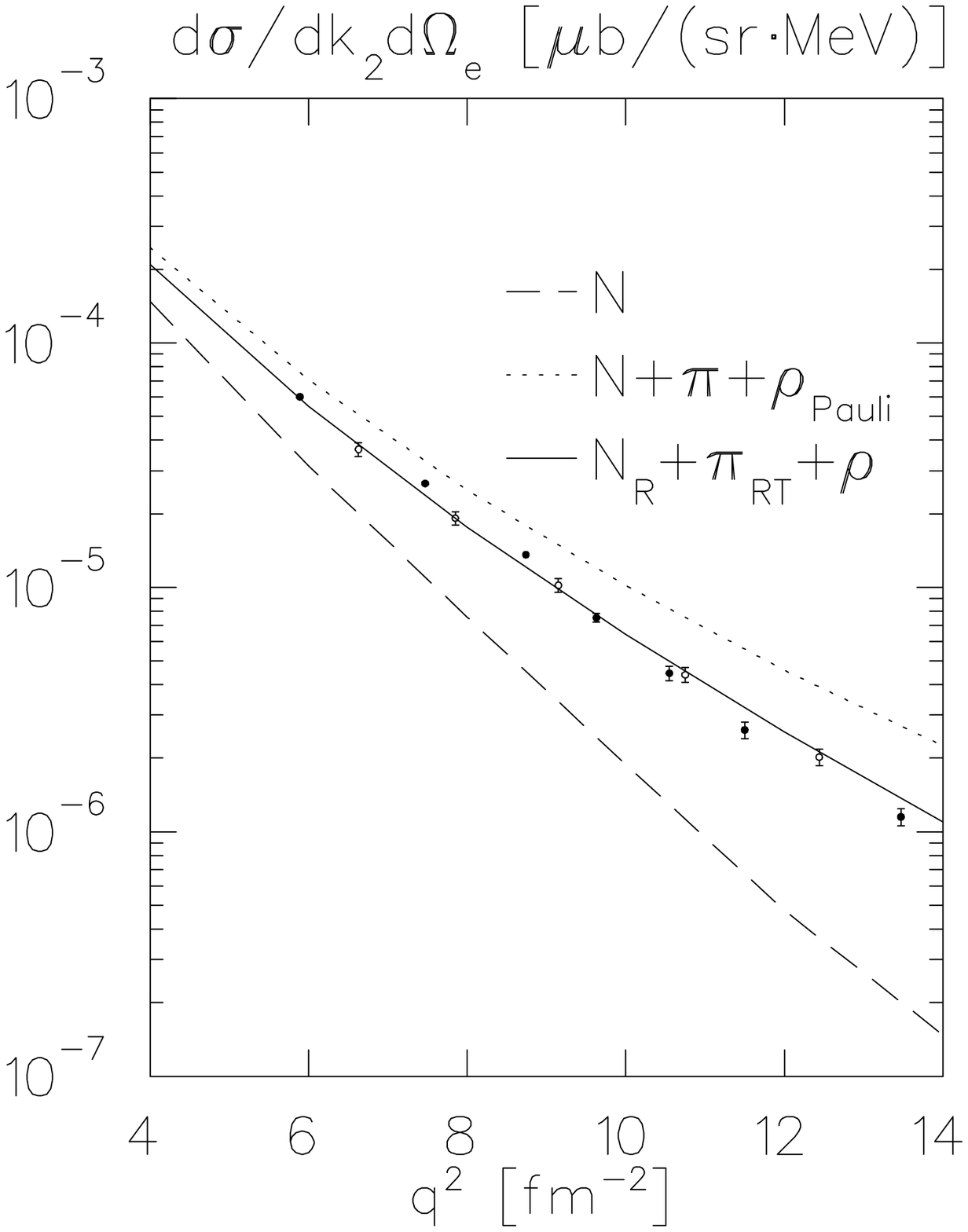,width=4.8cm}}
\noindent
{\small Fig.\ 2: Differential cross section for $d(e,e')np$ near threshold 
at backward angles. Dashed: normal theory without MEC; dotted: with 
nonrelativistic MEC; full: with RC (from~\cite{Rit95})}
\label{fig:th2}
\end{minipage}
\smallskip

First, I will consider the inclusive process $d(e,e')pn$, where no analysis
is made of the hadronic final state. The threshold region is dominated 
by the excitation
of the antibound $^1S_0$-resonance in NN scattering at
very low energies. It is the inverse process to thermal n-p radiative
capture, which proceeds via $M1$ transition and where MEC  and IC
give about a  10  percent  enhancement~\cite{RiB72,Are81b}.  The
advantage of electron scattering in having the momentum transfer at one's
disposition  becomes  now  apparent  since  the  relatively  small  effect  
of subnuclear d.o.f.\ in the real photon process can be amplified 
considerably. The reason for this lies in the
fact that with increasing momentum transfer the one-body contribution drops
rapidly due to a destructive interference of S- and D-wave contributions
and thus the distribution of the momentum transfer onto both nucleons via the 
two-body operators becomes more favourable. This can be seen in Fig.\ 1
where the transverse form factor is shown as obtained from the Rosenbluth 
separation~\cite{SiB79}. While the longitudinal form factor
is well described by the classical theory
supporting the Siegert hypothesis, that the charge density is little
affected by exchange effects,
one finds a large discrepancy in $F_T$ which is only resolved if MEC and IC
are added. 

The situation for higher momentum transfers is shown in Fig.\ 2 
where the inclusive cross section is plotted at backward angles for 
moderate momentum transfers. One readily sees the strong 
enhancement due to MEC, but in addition one notes sizeable relativistic 
contributions. 
It is clear from this result that already at low excitation energies 
relativistic effects may become important and have to be considered in a 
quantitative comparison of theory with experiment. One has to keep in mind 
that relativistic contributions appear both in the current operators and 
in the wave functions. For the latter case one may
distinguish between (i) the internal dynamics of the rest frame wave
function and (ii) the boost operation transforming the rest frame wave
function into a moving frame. It is obvious that for a conclusive 
interpretation one has to include all corrections of the same order 
consistently. In this work, presented in Fig.\ 2, where we have adopted 
a $(p/M)^2$-expansion starting from a covariant approach~\cite{ATA89}, 
all relativistic terms are included consistently. 

\smallskip
\noindent
\begin{minipage}[t]{5.5cm}
\centerline{\psfig{figure=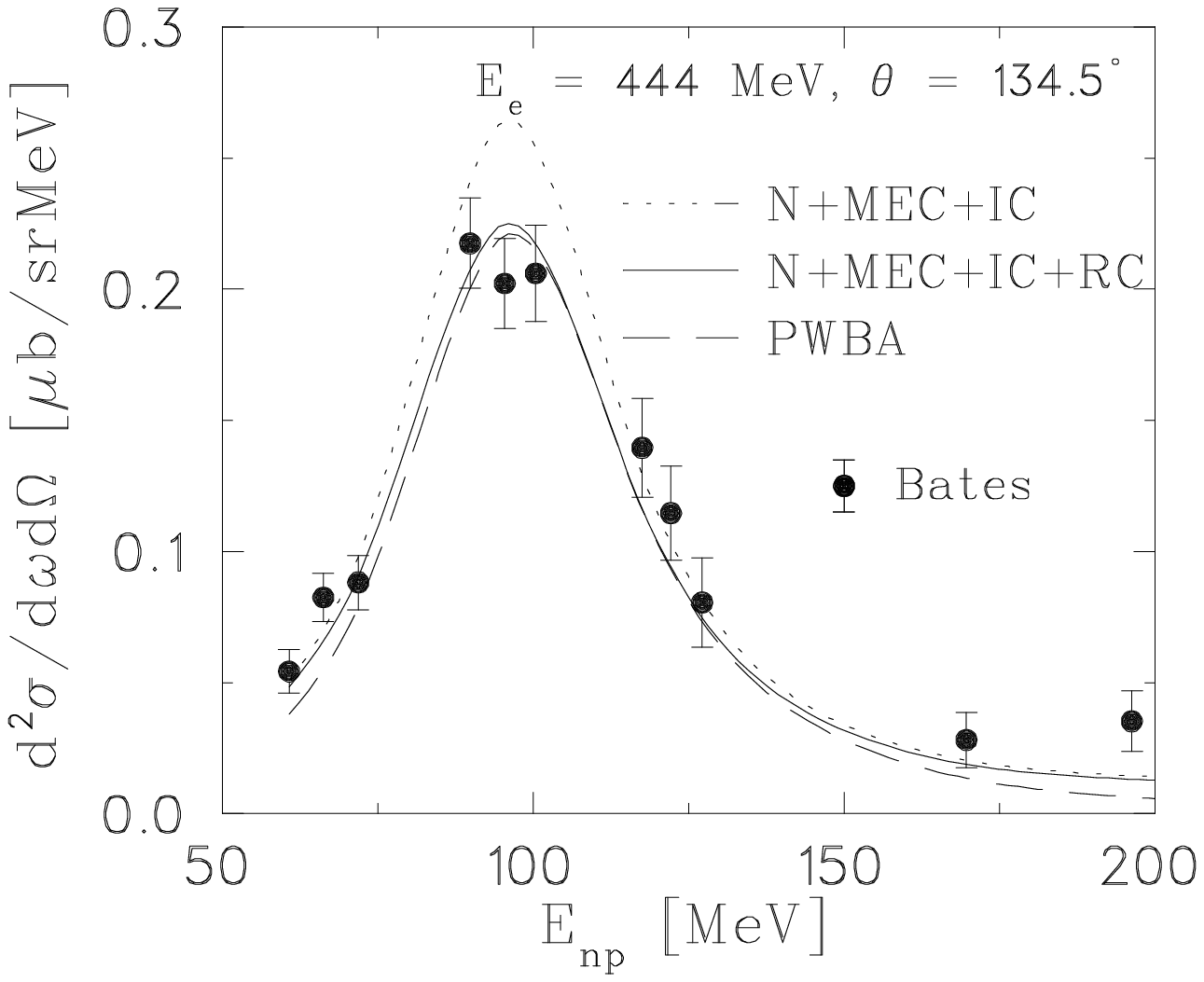,width=5.1cm,angle=90}}
\noindent
{\small Fig.\ 3: 
Inclusive cross section for $d(e,e')$ for PWBA and
with FSI, MEC, IC, and RC (from~\cite{WiB93}).} 
\label{fig:bates}
\end{minipage}\hfill
\begin{minipage}[t]{5.5cm}
\centerline{\psfig{figure=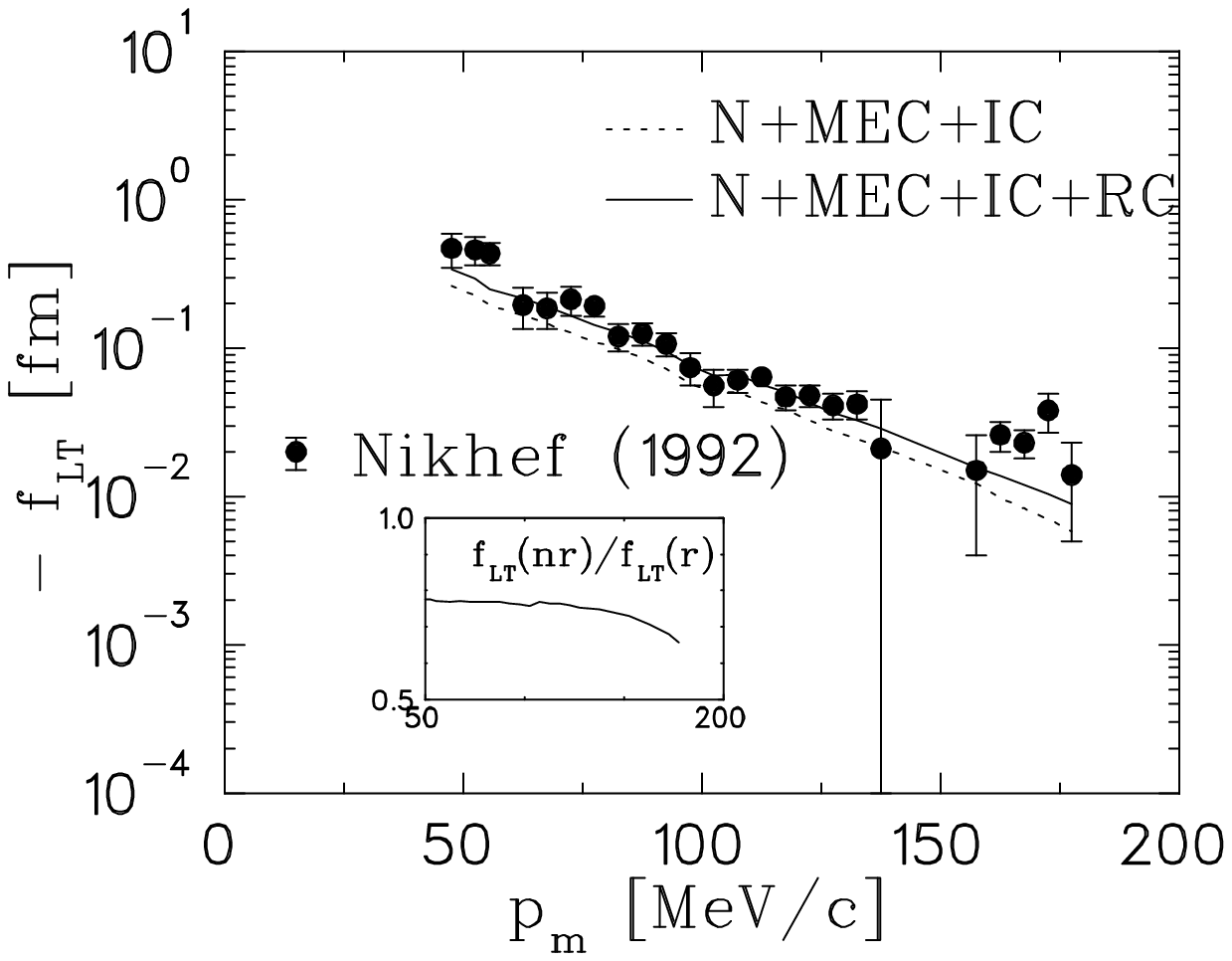,width=5.5cm,angle=90}}
\noindent
{\small Fig.\ 4: 
Coincidence structure function $f_{LT}$ 
as function of the missing momentum without and with
relativistic corrections (from~\cite{BeW94}).}
\label{fig:nikhef}
\end{minipage}

In Fig.\ 3 we show for higher energy and momentum transfer another comparison 
of experimental data for $d(e,e')$ with a realistic calculation including 
FSI, MEC, IC and the dominant relativistic contributions (RC) and a comparison 
to the pure plane wave Born approximation (PWBA). One notes that the 
relativistic contributions improve considerably
the agreement between theory and experiment. Furthermore, the comparison to
the PWBA shows that off the quasi-free region interaction effects are
important. 

As last example of relativistic effects, I show in Fig.\ 4 the 
longi\-tudinal-trans\-verse inter\-ference struc\-ture function $f_{LT}$ 
ap\-pearing in the ex\-clu\-sive process $d(e,e'p)n$.
The com\-parison with recent ex\-peri\-mental data 
is signifi\-cantly improved if relati\-vistic 
con\-tri\-butions are included. The relative size of the RC is shown in the 
inset as ratio of non\-relativistic (nr) to relativistic (r) result. 

\subsection{Signature of a $\Delta\Delta$ component in the deuteron}

\noindent
\begin{minipage}[t]{5.5cm}
\centerline{\psfig{figure=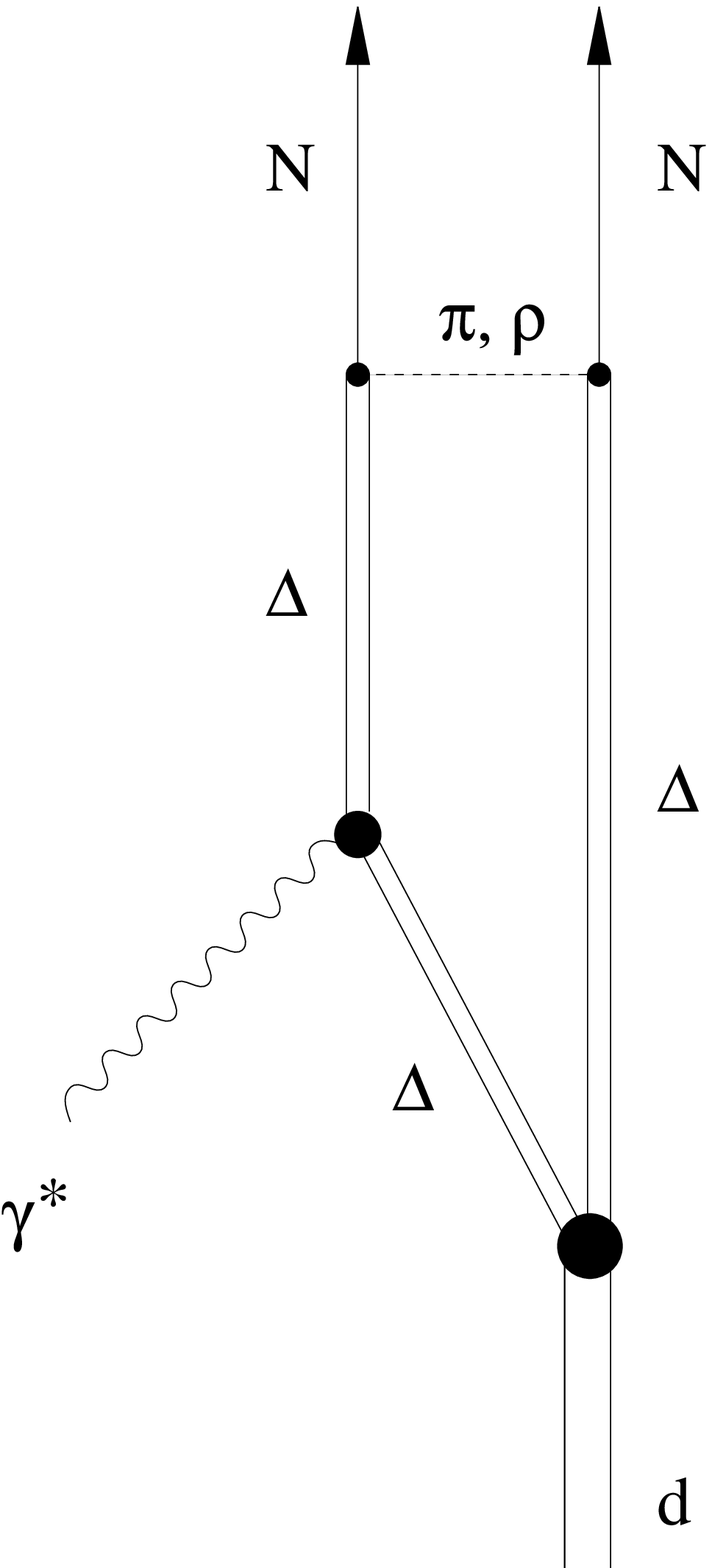,height=4cm}}
\noindent
{\small Fig.\ 5: Contribution of the $\Delta\Delta$ component to the 
longitudinal structure function $f_{L}$}
\label{fig:dd_component}
\end{minipage}\hfill
\begin{minipage}[t]{6cm}
\centerline{\psfig{figure=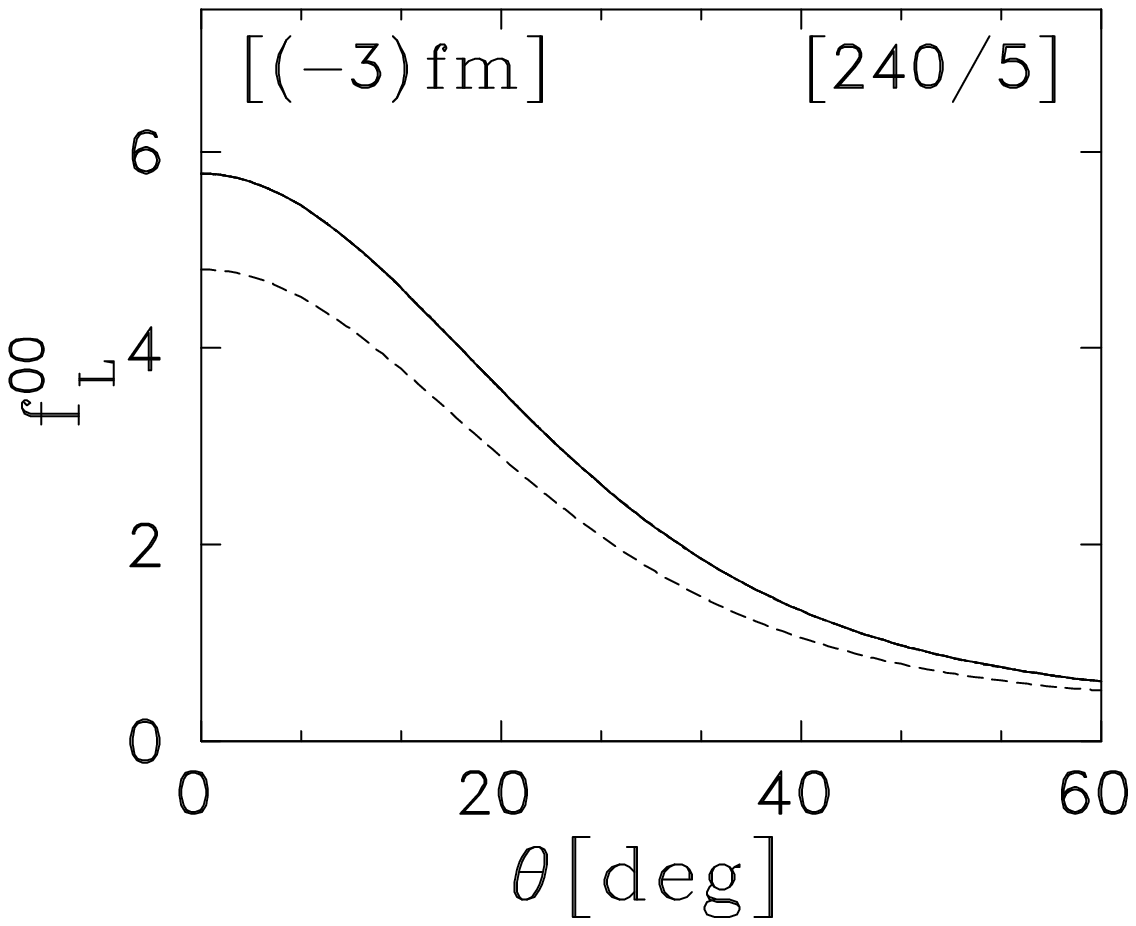,width=5cm}}
\noindent
{\small Fig.\ 6: Longitudinal structure function in the $\Delta$ region for 
$E_{np}=240$ MeV and $\vec q^{\,2}=5$ fm$^{_2}$ without (dashed) and with 
(full) $\Delta\Delta$ component.}
\label{fig:dd}
\end{minipage}
\smallskip

One important consequence of the internal nucleon dynamics is the presence 
of small wave function components where one or more nucleons are internally 
excited into an isobar state~\cite{WeA78}. 
These isobar configurations (IC) are rather 
small and thus their presence is difficult to detect. In any case, evidence 
for them will only be indirect since they are not directly observable. 
However, under certain favourable conditions they may lead to sizeable 
effects. I would like to show one example for the double $\Delta$-component 
in the deuteron. 

In view of the fact, that the charge excitation of a $\Delta$ which has to 
proceed via $C2$ is largely suppressed, the only contribution of IC to the 
longitudinal structure function $f_{L}$ comes from the double 
$\Delta$-component as sketched in the diagram of Fig.\ 5. At low momentum 
transfer, the contribution is negligible. However, with increasing momentum 
transfer its relative importance is enhanced considerably because of the 
much shorter ranged structure of the isobar configurations compared to the 
normal ones. This behavior is shown in Fig.\ 6. It also illustrates nicely the 
advantage of varying the momentum transfer independently in electron 
scattering.

\subsection{$N\Delta$ dynamics in $d(\gamma,p)n$ in a coupled channel 
approach}
Deuteron photodisintegration is fairly well understood at low energies
in the framework of nucleon, meson and isobar
d.o.f.~\cite{SWA 91} whereas at higher photon energies
between $200$ and $450\,$MeV, where the $\Delta$-excitation dominates
the reaction, the theoretical description is much less well
settled~\cite{ArS 91}. Often the $\Delta$-excitation is treated in the impulse 
approximation (IA). It turns out, however, that it is important to include 
the $N\Delta$ dynamics in a coupled channel approach~\cite{LeA 87,TaO 89}.
Here I will present an improved coupled $NN$-$N\Delta$ channel approach 
which includes in addition explicit pion d.o.f.~\cite{WiA 93}. In this 
calculation, the  
dominant magnetic dipole excitation of the $\Delta$ 
has been fitted to the experimental 
$M_{1+}(3/2)$ multipole of pion photoproduction on the nucleon. 
Details can be found in Ref.~\cite{WiA 93}. 

\smallskip
\noindent
\begin{minipage}[b]{4cm}
\noindent
{\small Fig.\ 7: Total cross section for $\gamma d\rightarrow pn$ in 
comparison with
experiment. 
Complete calculation (full), IA (dotted) and complete calculation
with modified $\gamma N\Delta$-coupling (dashed) as described in the text 
(from~\cite{WiA 93}).
}
\label{fig:dgtot}
\end{minipage}\hfill
\begin{minipage}[t]{7.5cm}
\centerline{\psfig{figure=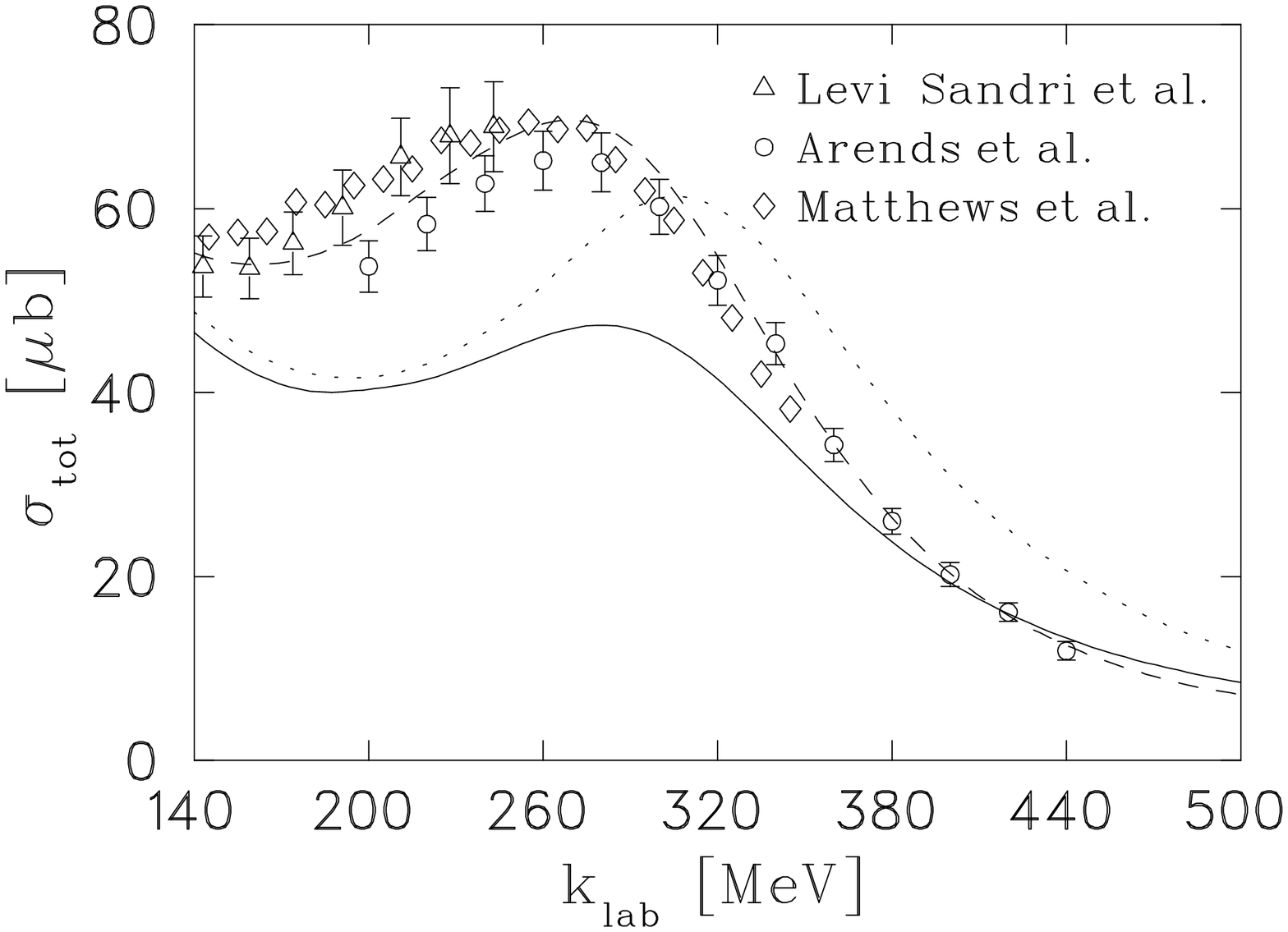,width=7cm,angle=90}}
\end{minipage}

I first show in Fig.\ 7 the total cross section. 
In comparison with the IA, the full calculation leads to a strong
reduction of the cross section above $260\,$MeV
and in addition to a shift of the maximum position
towards lower energies in accordance with the experimental
energy dependence,
whereas the IA peaks close to the resonance position of
the free $N\Delta$ system at $320\,$MeV. 
However, below $340\,$MeV the cross section becomes definitely too small.
Since the $\gamma N\Delta$-coupling is weaker than the effective one used 
in~\cite{LeA 87} 
we have also considered a modified $\gamma N\Delta$-coupling,
which was determined from the elementary amplitude
under the assumption of vanishing nonresonant
contributions to the $M_{1+}(3/2)$ multipole. Using this coupling, we achieved
a good agreement of the total cross section with experiment
over the whole energy range as demonstrated in Fig.\ 7. 
Since in this case the Born terms are effectively incorporated in the 
modified coupling we are led to the conclusion that
the framework of static $\pi$-exchange currents, 
containing in principle the Born terms,
gives a poor description of them in this energy region.

\smallskip
\centerline{\psfig{figure=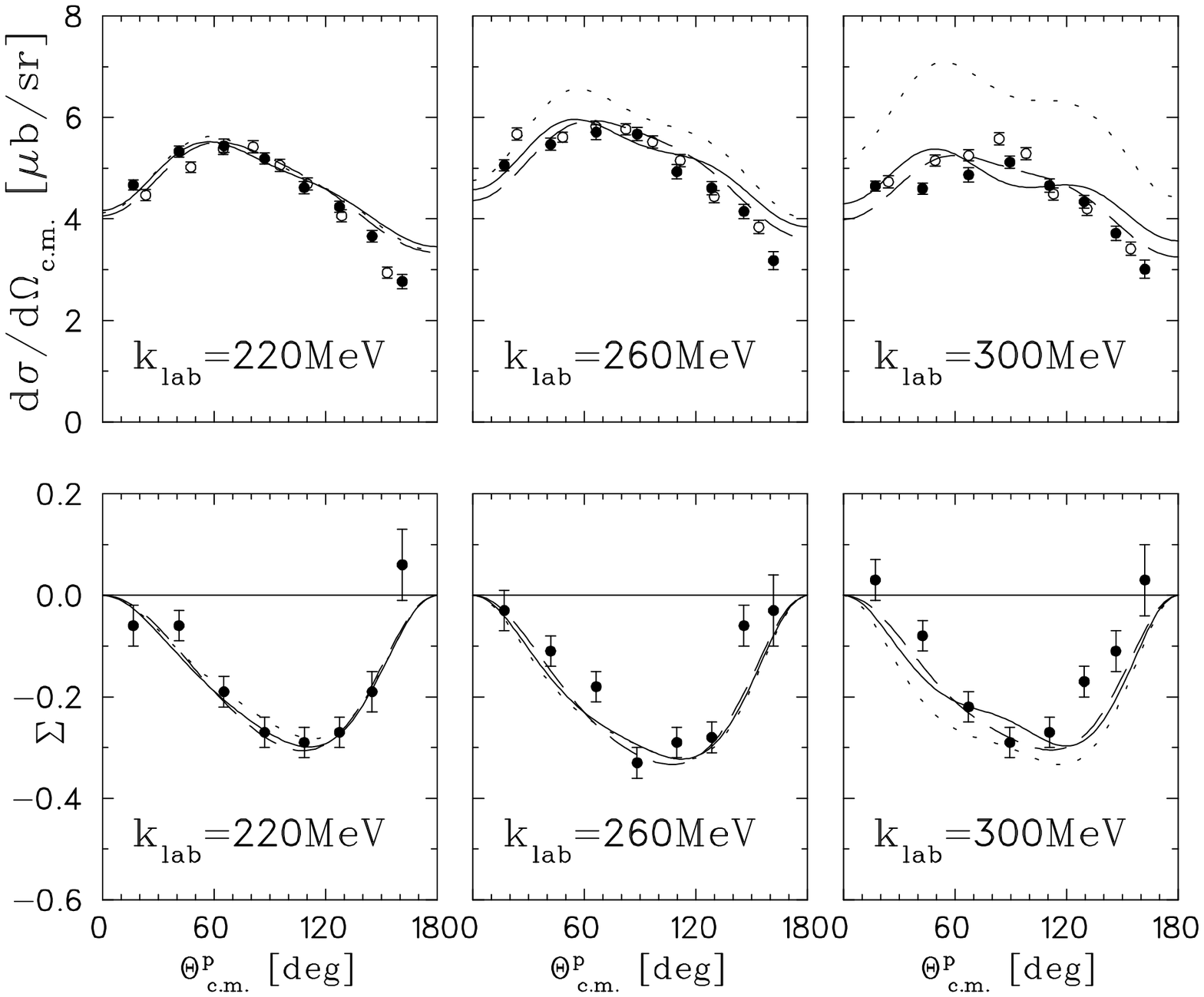,width=8cm}}
\noindent
{\small Fig.\ 8: Differential cross section 
and photon asymmetry $\Sigma$ 
for three energies:
complete calculation with modified $\gamma N\Delta$-coupling (full),
IA (dotted) and complete calculation with modified $\gamma
N\Delta$-coupling in M1 only (dashed) (from~\cite{WiA 93}).}
\label{fig:dgnp}

Differential cross sections and photon asymmetries for three energies are 
shown in Fig.\ 8. In addition to the IA and the
complete calculation with modified coupling, we show results where the
$\gamma N\Delta$-coupling has been modified in the M1 multipole only.
Despite the good description of the total cross section, problems in the angular
distributions still remain.
At the lower energy our calculation does not show such a strong decrease
at the backward angles as the data.
Furthermore, a dip around $90^{\circ}$ appears, in particular at the 
highest energy, which clearly contradicts the experimental shape.
The occurrence of this dip structure is also reported in~\cite{TaO 89}.
The photon asymmetry in Fig.\ 8 is quite satisfactory although 
at the higher energies it appears to be more asymmetrical around $90^\circ$ 
than the experiment. At even higher energies this trend continues so that the 
model clearly fails to reproduce the data.

\subsection{Unitary ambiguity in the resonance multipoles for 
$\gamma + N \rightarrow \Delta$}

From the foregoing it is clear that for a reliable description of $\Delta$
excitation in nuclei one needs to know their multipoles fairly well. 
Consequently, there is considerable experimental effort in
measuring the corresponding $E_{1+}$ and $M_{1+}$ isospin $3/2$
multipole amplitudes for photoproduction of pions on the 
nucleon.
However, all realistic pion photoproduction models
show that both multipoles, in particular $E_{1+}^{3/2}$, contain
nonnegligible nonresonant background contributions. Unfortunately,
their presence complicates the isolation of the resonant parts.
Basically two different approaches are used in
order to extract these transition
amplitudes.  The first one is an effective lagrangian approach, 
in which the $\pi N$ scattering is not treated dynamically and thus 
different unitarization schemes are used. These introduce some model 
dependence. 

\smallskip
\centerline{\psfig{figure=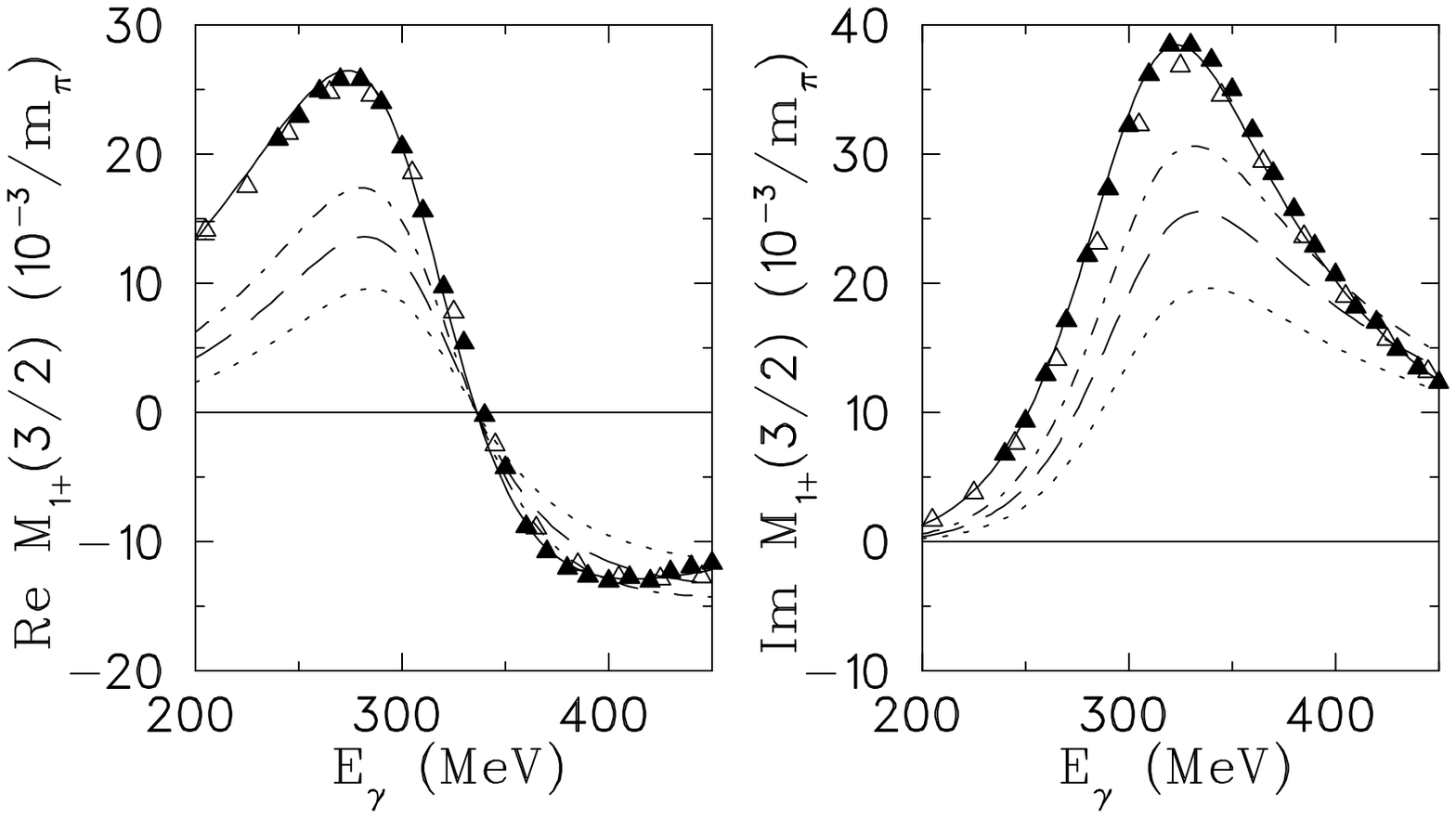,width=8cm}}
\noindent
{\small Fig.\ 9: Real and imaginary parts of the $M_{1+}^{3/2}$ multipole 
as function of the photon laboratory energy $E_\gamma$.
Dashed, dotted and dash-dotted curves are bare multipoles
corresponding to transformation angles $\tilde \alpha = 0^\circ$,
$10^\circ$ and $-10^\circ$, respectively. The solid curves show the total
multipoles which are representation independent (from~\cite{WiW96}).}
\label{fig:multipol}

In the second 
approach, the $\pi N$ interaction is treated dynamically and thus
unitarity is respected automatically. 
However, also the latter approach contains considerable model dependence of the 
resonance properties because of an inherent unitary ambiguity. 
The reason for this unitary freedom is that the separation of a resonant 
$\Delta$
contribution corresponds to the introduction of a $\Delta$ component
into the $\pi N$ scattering state which vanishes in the asymptotic region. 
It is known that the explicit form of a wave function depends on the chosen
representation, and can be changed by means of unitary
transformations. Consequently, the probability of a certain wave
function component is not an observable, since it depends on the
representation, whereas any observable is not affected by a change of the 
representation. 
We have demonstrated this fact in a simple model where the unitary 
transformation mixes resonant and background $\pi N$ interactions~\cite{WiW96}. 
I show in Fig.\ 9 the effect of such unitary transformations (governed by a 
parameter $\tilde \alpha$) on the $M_{1+}^{3/2}$ multipole. One notices a 
considerable variation of the bare resonance multipole with $\tilde \alpha$ 
while the total one is unchanged.


With this I would like to close the brief review in which I could give 
only a very cursory survey on the many interesting facets of present day 
research in current electromagnetic physics programs at various 
c.w.\ electron machines. 

\section*{Acknowledgments}
The collaboration of J.\ Adam, G.\ Beck, H.\ G\"oller, F.\ Ritz, 
Th.\ Wilbois and P.\ Wilhelm is gratefully acknowledged. This work has been 
supported by the Deutsche Forschungsgemeinschaft (SFB 201). 



\end{document}